\documentclass[reprint,aps,pra,superscriptaddress]{revtex4-1}

\usepackage{graphicx}
\usepackage{amsmath}
\usepackage{amssymb}
\usepackage{dcolumn}
\usepackage{bm}
\usepackage{setspace}
\usepackage{color}

\begin{document}

\title{\LARGE Orientation control and nonlinear trajectory tracking of colloidal particles using microfluidics}

\author{Dinesh Kumar}
\thanks{These authors contributed equally}
\affiliation
{
  Department of Chemical and Biomolecular Engineering \\ University of Illinois at Urbana-Champaign, Urbana, IL, 61801
}
\affiliation
{
Beckman Institute for Advanced Science and Technology \\ University of Illinois at Urbana-Champaign, Urbana, IL, 61801
}

\author{Anish Shenoy}
\thanks{These authors contributed equally}
\affiliation
{
  Department of Mechanical Science and Engineering \\ University of Illinois at Urbana-Champaign, Urbana, IL, 61801
}
\author{Songsong Li}
\affiliation
{
  Department of Materials Science and Engineering \\ University of Illinois at Urbana-Champaign, Urbana, IL, 61801
}

\author{Charles M. Schroeder}
\email[To whom correspondence must be addressed: ]{cms@illinois.edu}
\affiliation
{
  Department of Chemical and Biomolecular Engineering \\ University of Illinois at Urbana-Champaign, Urbana, IL, 61801
}
\affiliation
{
Beckman Institute for Advanced Science and Technology \\ University of Illinois at Urbana-Champaign, Urbana, IL, 61801
}
\affiliation
{
  Department of Materials Science and Engineering \\ University of Illinois at Urbana-Champaign, Urbana, IL, 61801
}

\date{\today}

\begin{abstract}

Suspensions of anisotropic Brownian particles are commonly encountered in a wide array of applications such as drug delivery and manufacturing of fiber-reinforced composites. Technological applications and fundamental studies of small anisotropic particles critically require precise control of particle orientation over defined trajectories and paths. In this work, we demonstrate robust control over the two-dimensional (2D) center-of-mass position and orientation of anisotropic Brownian particles using only fluid flow. We implement a path-following model predictive control scheme to manipulate colloidal particles over defined trajectories in position space, where the speed of movement along the path is a degree of freedom in the controller design. We further explore how the external flow field affects the orientation dynamics of anisotropic particles in steady and transient extensional flow using a combination of experiments and analytical modeling. Overall, this technique offers new avenues for fundamental studies of anisotropic colloidal particles using only fluid flow, without the need for external electric or optical fields. 

\end{abstract}

\maketitle

Anisotropic particles play an integral role in the scalable fabrication of mesostructured composite materials \cite{advani_molding_2012}. Such composites are routinely processed in complex flows described by a combination of shear and elongation. Achieving a quantitative understanding of how fluid flow orients particles and impacts the rheological behavior of anisotropic particle and fiber suspensions demands precise tracking and control of particle position and orientation. To this end, significant efforts have been devoted to develop methods for precisely controlling the motion of single and multiple anisotropic colloidal particles \cite{ashkin_optical_1987,galajda_orientation_2003,yu_manipulation_2004,tong_alignment_2009,bingelyte_optically_2003,mathai_simultaneous_2011}. In particular, methods based on optical tweezers \cite{galajda_orientation_2003,yu_manipulation_2004,tong_alignment_2009,kang_angular_2012} and electrokinetic traps \cite{mathai_simultaneous_2013} have been used for the simultaneous manipulation of position and orientation of anisotropic particles. Such methods hold the potential to directly benefit the field of directed assembly, which aims to precisely assemble chemically and structurally distinct anisotropic particles into functional hierarchical structures. Towards this goal, holographic optical tweezers have been used for multiplexed trapping of large ensembles of particles \cite{chiou_massively_2005} and for the directed assembly of dielectric rod-like particles into desired structures by effectively controlling their translation and orientation \cite{marago_optical_2013}. Electrophoretic and dielectrophoretic forces have also been used to manipulate the trajectory and orientation of small particles such as cytokine-conjugated nanowires, which was used to deliver molecular doses of biologically active chemicals to a specific site in a cell \cite{fan_precision_2008,fan_subcellular-resolution_2010}. In all cases, methods based on optical traps \cite{ashkin_optical_1987}, magnetic tweezers \cite{gosse_magnetic_2002}, and electrokinetic traps \cite{cohen_suppressing_2006,armani_using_2006} require application of an external field and generally rely on exploiting the intrinsic properties of the target particle or surrounding medium (e.g. magnetic susceptibility, polarizability, charge, or medium conductivity) for controlling the motion of individual particles. However, it is desirable to control the motion of small particles using methods that are independent of material composition. 

Hydrodynamic trapping offers an alternative method for controlling small particles that relies only on fluid flow \cite{shenoy_stokes_2016}. Hydrodynamic traps confine particles via frictional forces imposed by a flowing fluid, which generally poses no constraints on the chemical properties or material composition of trapped particles. Hydrodynamic forces were first used to trap large millimeter-sized droplets by G. I. Taylor, who developed a four-roll mill apparatus to generate mixed flows that can be varied from purely rotational to extensional flow \cite{taylor_formation_1934}. In 1985, Bentley and Leal \cite{bentley_computer-controlled_1986} developed a computer-controlled four-roll mill that allowed for controlling the position of millimeter or micron-sized particles near a stagnation point for extended periods of time. The advent of microfluidics has enabled several researchers to build microfluidic analogs of the four-roll mill \cite{hudson_microfluidic_2004,lee_microfluidic_2007}, though these methods do not explicitly incorporate automated feedback to control particle position or residence time in flow. Recently, a feedback-controlled hydrodynamic trap was developed for automated trapping of micro- and nanoscale particles in a PDMS-based microfluidic device equipped with an on-chip metering valve \cite{tanyeri_microfluidic-based_2011,tanyeri_manipulation_2013,shenoy_characterizing_2015}. These initial studies used a combination of proportional, integral and derivative control schemes for manipulating the two-dimensional (2D) center-of-mass position of spherical particles. Schroeder and coworkers further developed a multiplexed technique for controlling the 2D center-of-mass position of multiple particles known as the Stokes trap \cite{shenoy_stokes_2016}. The Stokes trap relies on a model predictive control scheme to independently manipulate single or multiple particles in solution along arbitrary trajectories by the sole action of fluid flow. Despite recent progress, however, automated hydrodynamic trapping methods have not been used to simultaneously control the orientation and position of anisotropic particles. However, the ability to precisely control the motion and alignment of single or multiple non-spherical particles in flow would greatly benefit several fields of research ranging from fundamental fluid mechanics \cite{Furst2017,Bechtel2018}, particulate flows, colloidal suspensions \cite{Swan2014,Leahy2017}, and directed assembly of materials. 

\begin{figure}[t]
\small
\centering
\includegraphics[width=0.5\textwidth]{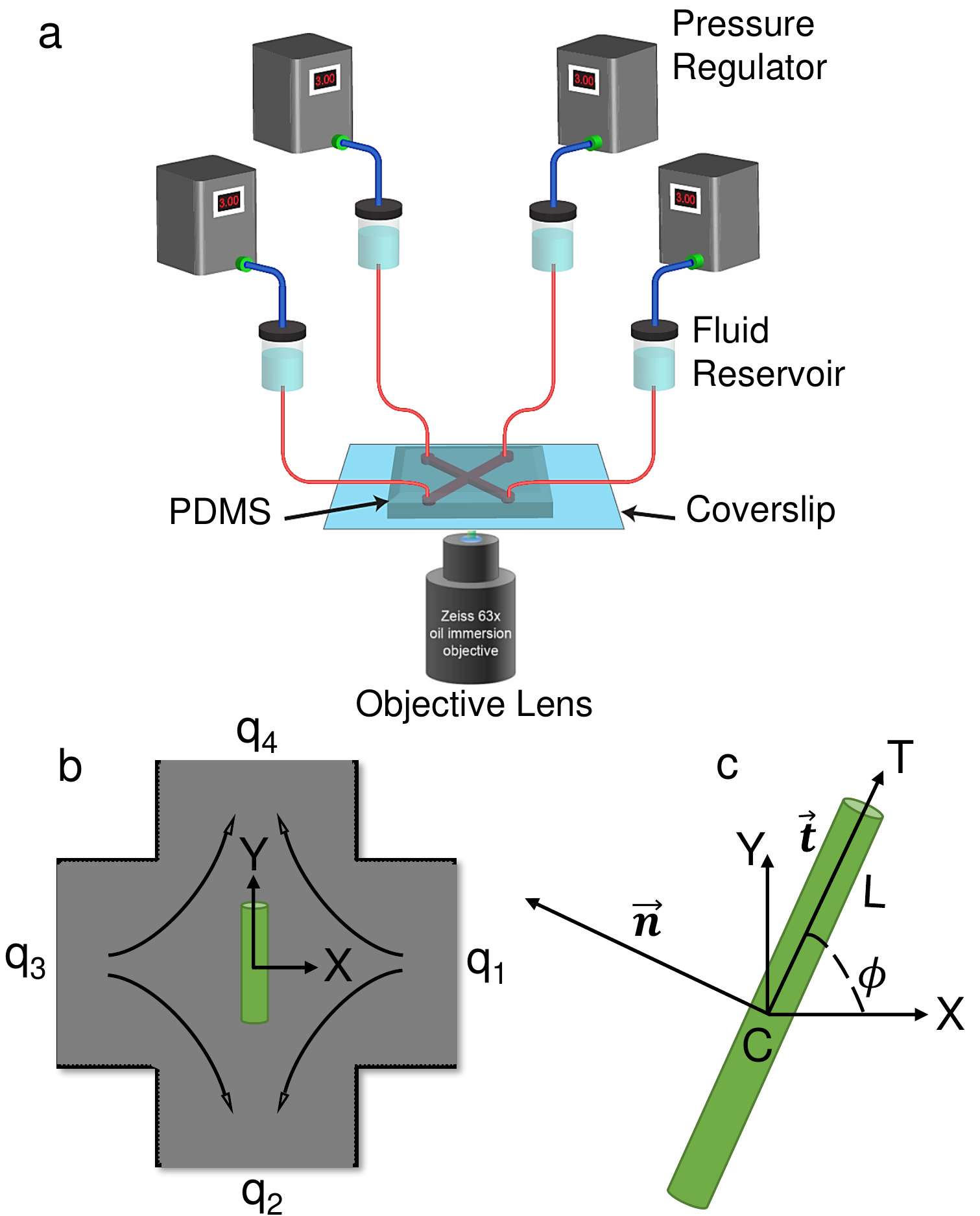}
\caption{\label{fig:rod_device} Overview of trapping method and anisotropic particle orientation. (a) Schematic of the experimental setup. Inlet/outlet ports of the microfluidic device are connected to fluidic reservoirs that are pressurized by regulators controlled by a custom LabVIEW code. (b) Schematic showing the top view of $\textit{M}=4$ channel microfluidic device for manipulating a single anisotropic particle at the center of slot. (c) Schematic of the orientation angle $\phi$ of an anisotropic particle, which is measured with respect to $x$-axis as shown in Fig.\ref{fig:rod_device}b in anti-clockwise direction. The tangent vector $\bm{t}$ along a line connecting the points $C$ and $T$ and the normal vector $\bm{n}$ define the particle orientation. The rod half-length is \em{L}. }
\end{figure}

In this work, we present two major developments in flow based particle trapping. First, we demonstrate simultaneous flow-based control of the orientation and center-of-mass (COM) position of an anisotropic Brownian particle. We quantify the trapping performance using this new method and determine the translational and angular trap stiffness which depend on the flow strength and the directions of the principal axes of compression and extension. We use this method to directly observe the transient and steady state dynamics of a single anisotropic Brownian particle in extensional flow over long times and compare the experimental results with analytical models. Second, moving beyond simple set-point stabilization, we implement a non-linear path following algorithm that substantially improves the accuracy and the speed with which arbitrary smooth paths can be tracked by particles. Together, these advances will facilitate fundamental studies of colloidal particles requiring precise control over COM position and orientation, as well as control of geometrical path trajectories over long times. 

\begin{figure*}
\small
\centering	
\includegraphics[width=1.0\textwidth]{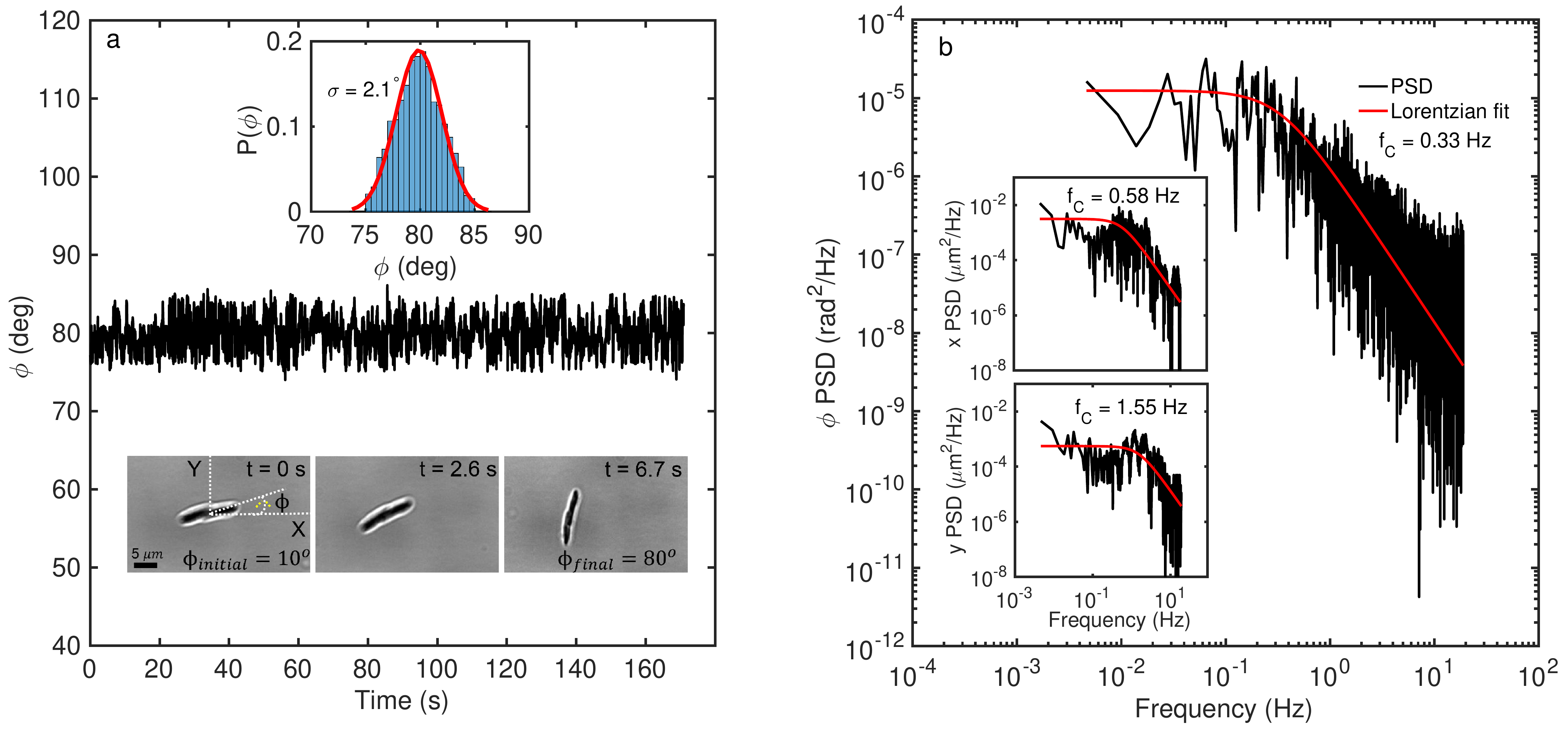}
\caption{\label{fig:trap_stiff} Simultaneous control over position and orientation of anisotropic colloidal particles. (a) Trajectory of angular displacements of a trapped Brownian rod over a period of 170 s. (\textit{Inset, Upper}) Probability distribution function of particle orientation; Gaussian fit shown in red. The standard deviation of rotational displacement $\sigma$ is equal to $2.1^{\circ}$. (\textit{Inset, Lower}) Sequential images of a trapped Brownian rod while controlling the COM position and orientation. At $t=0$, the rod is initially oriented at $\phi=10^{\circ}$ with respect to the $x$-axis. At $t=0.033$  s, the target orientation is changed to  $\phi=80^{\circ}$. The final desired state is achieved within 6.7 s, yielding a rod average angular velocity of $\approx$ 0.16 rad/s.  (b) Power spectral density of orientation angle fluctuations for the trajectory shown in Fig. \ref{fig:trap_stiff}a. The PSD is fit to a Lorentzian function, enabling determination of the corner frequency $f_c$ along the $\phi$ direction. (\textit{Inset, Upper}) Power spectrum of the $x$-position translational displacement acquired in the Fourier domain with Lorentzian fitting. The corner frequency is 0.58 Hz in the $x$-axis. (\textit{Inset, Lower}) SPower spectrum of the $y$-position translational displacement acquired in the Fourier domain with Lorentzian fitting. The corner frequency is 0.58 Hz in the $y$-axis.}
\end{figure*}

\section{Results}
\subsection{Flow model and governing equations}
The governing principles of the trapping method can be understood by considering a general problem of trapping \textit{P} particles independently in a microfluidic device with \textit{M} channels intersecting at an angle of $2\pi/M$ to form an \textit{M}-sided polygonal device \cite{schneider_algorithm_2011}. Fig. \ref{fig:rod_device}a,b shows a schematic of a microfluidic device with $M=4$ channels. The objective is to manipulate the 2D center-of-mass position and orientation of $P$ anisotropic particles (Supplemental Material and Fig. S1), which involves controlling 3\textit{P} degrees of freedom (states) concurrently. The control inputs for this problem are the transient incoming/outgoing fluid streams through the \textit{M} channels. However, the flow incompressibility condition reduces the number of independent control inputs to $M-1$. To effectively control the system, the number of independent control variables should at least be equal to the number of states being controlled such that $M\geq3P+1$. Therefore, a microfluidic device with $M=4$ channels provides a sufficient platform (Supplemental Material and Fig. S2) for manipulating the center-of-mass position and orientation of a single anisotropic particle.

We first consider the fluid dynamics inside the microfluidic device. In Stokes flow, the inertial terms in the Navier-Stokes equation can be neglected and we can write \cite{leal_advanced_2007}:
\begin{equation}\label{eq:stokesflow} 
-\nabla p +\mu\nabla{}^2 \bm{u} = \bm{ 0 } 
\end{equation}
where $p$ is the dynamic pressure field, $\bm{u}$ is the velocity field, and $\mu$ is the viscosity. In a microfluidic cross-slot geometry generated by the intersection of $M$ channels, we can approximate each channel as a point source of flow \cite{shenoy_stokes_2016}. The height-averaged velocity field at a point $\bm{x}$ inside the cross-slot can then be approximated as a linear superposition of the velocity fields generated by each point source such that:
\begin{equation}\label{eq:pointsource}
\bm{u} = \frac{1}{\pi  H} \sum^M_{i=1} \frac{(\bm{x}-\bm{R}^i)q_i}{\|\bm{x}-\bm{R}^i\|^2}  \triangleq \bm{F}(\bm{x},\bm{q},\bm{R})
\end{equation}
Here, $H$ is the height of the device, $\bm{x}\in \mathbb{R}^2$ is a position vector of a point in flow, $\bm{R}^i \in \mathbb{R}^2$ is the the position vector of the $i^{th}$ point source, and $\bm{q}\in\mathbb{R}^M$ is a vector whose $i^{th}$ element $q_i$ represents the volumetric flow rate through the $i^{th}$ point source. The flow rates $q_i$ are not unconstrained and must satisfy mass conservation, which yields:
\begin{equation}\label{eq:sumflowrates}
\sum^M_{i=1}q_i = 0
\end{equation}
where flow rates are defined to be positive (negative) when they flow into (out of) the cross-slot. The orientation of anisotropic particles is modeled by considering two points along a rod-like object (Fig. \ref{fig:rod_device}c), with point $C$ located at the particle center-of-mass at $\boldsymbol{x}_c$ and a second point $T$ located at the rod terminus at $\boldsymbol{x}_t$. In this way, the unit tangent vector $\bm{t}$ along the rod is given by $\bm{t}=(\bm{x}_t - \bm{x}_c)/\left \| (\bm{x}_t-\bm{x}_c) \right \|$. Here, the length of the particle along the major axis is $2L$, where $L \equiv \left \| (\bm{x}_t-\bm{x}_c) \right \|$. Rod-like particles are defined by an orientation angle $\phi$ in the 2D plane given by the angle between the $x$-axis and the tangent vector $\bm{t}$. The scalar rotational velocity of an anisotropic particle $\dot{\phi}$ is given by the dot product of the relative fluid velocity vector $\bm{v}_{rel}$ between points $C$ and $T$ with the normal vector $\bm{n}$ divided by $L$, such that $\dot{\phi} = \left \| \bm{t} \times \bm{v}_{rel} \right \| / L = \bm{v}_{rel} \cdot \bm{n} / L$. This equation can be recast in terms of the particle center-of-mass position $\bm{x}_c$ and simplified to yield the rotational velocity (Supplemental Material):
\begin{equation}\label{eq:angularvelocity}
\dot{\phi} = \frac{1}{\pi H}\sum_{i=1}^{M}\frac{-2\bm{t}^{T}(\bm{x}_c-\bm{R}^i)(\bm{x}_c-\bm{R}^i)^{T}\bm{n}}{\left \| (\bm{x_c}-\bm{R}^i) \right \|^4}q_i \triangleq \bm{G}(\bm{x},\bm{q},\bm{R})
\end{equation}
where $\bm{n} = [-\sin \phi, \cos \phi]^T$ is the unit normal vector perpendicular to $\bm{t} = [\cos \phi, \sin \phi]^T$. Eqns. (\ref{eq:pointsource}), (\ref{eq:sumflowrates}), and (\ref{eq:angularvelocity}) completely determine the 2D velocity field and rotational velocity of an anisotropic particle in the cross-slot geometry. In the absence of external forces, we can assume that the particle center-of-mass moves with the local fluid velocity and that the particle rotates with the fluid. We further neglect hydrodynamic interactions and perturbations to the base flow field due to the presence of the finite-sized particles. The center of mass coordinates and the orientation of the particle are represented by the state vector $\bm{X} \triangleq  [\bm{x}_c,\ \phi]^T$. In this case, we can write the prediction model for a particle's linear and angular motion using the following equation:
\begin{equation}\label{eq:flowmodel}
\begin{aligned}
\dot{\bm{X}} = \bm{H}(\bm{X},\bm{q},\bm{R}) \triangleq 
\begin{bmatrix} 
\bm{F}(\bm{x}_c,\bm{q},\bm{R})\\
\bm{G}(\bm{x}_c,\bm{q},\bm{R})
\end{bmatrix}
\end{aligned}
\end{equation}

\subsection{Model predictive control scheme for set-point stabilization}
The objective is to manipulate a single anisotropic particle in real time from an initial state to a final state defined by a desired position and orientation. To achieve this goal, we use the superposed point source linear and angular velocity field given by Eq. \ref{eq:flowmodel}. However, this objective poses several practical challenges for experimental implementation. First, we seek to manipulate micron- or sub-micron-sized particles that are subject to Brownian motion and thus follow non-deterministic trajectories. In addition, the fluidic model described in Eq. \ref{eq:pointsource} is merely an approximate model based on point sources, and any control strategy must be sufficiently robust to handle model imperfections. Finally, the particle state (position and orientation) is sampled at a rate of 30 Hz, hence the control strategy should be capable of calculating the optimal control scheme within 33 ms. For these reasons, we use model predictive control (MPC) to precisely manipulate particles \cite{mayne_constrained_2000, shenoy_stokes_2016}.  

Consider the task of manipulating a single particle from an initial state $\bm{X}_0$ to a final state $\bm{X}_F$, where the state is defined by both COM position and orientation. Although there are infinitely many trajectories in positional and orientational space between these states, we would like to select a trajectory that simultaneously minimizes flow rates as well as the translational and angular distance traveled. We can systematically obtain these trajectories and the corresponding flow rates at each sampling instance by minimizing the objective function $J$:
\begin{subequations}\label{eq:singleparticlecontrol}
	\begin{align}
  &	\underset{\tilde{\bm{X}}, \tilde{\bm{q}}}{\min }\ J  \nonumber\\
  & =\sum^{t_k+T_N-1}_{\tau=t_k} 
  \left\{ 
  \left(\tilde{\bm{X}}(\tau)-\bm{X}_F \right)^T \bm{\alpha} 
  \left(\tilde{\bm{X}}(\tau)-\bm{X}_F \right)
  +  \beta\tilde{\bm{q}}^T \tilde{\bm{q}}   \right\} \nonumber \\
& +  \gamma \ (\tilde{\bm{X}}(t_k+T_N) - \bm{X}_F)^T(\tilde{\bm{X}}(t_k+T_N) - \bm{X}_F) \\ 
&	\text{s.t.} \quad \frac{d \tilde{\bm{X}}}{dt}= \bm{H}(\tilde{\bm{X}},\tilde{\bm{q}},\bm{R}),\quad \tilde{\bm{X}}(t_k) = \bm{X}(t_k) \label{eq:mpcsysmodel}\\
&	\sum^N_{i=1}   \tilde{q}_i(\tau) =0 \quad \forall \ \tau = t_k,\ldots, t_k+T_N
	\end{align}
\end{subequations}
where $t_k = t_0 + k\Delta$ represents the $k^{th}$ sampling instant and $T_N$ is the MPC horizon, consisting of $N$ regular intervals such that $T_N = N\Delta$, where $\Delta$ is the sampling interval. $\tilde{\bm{X}}(\tau)$ is the predicted value of the position and orientation at time $\tau$, and $\tilde{\bm{q}}(\tau)$ indicates the predicted piecewise constant control applied during the interval $[\tau,\tau+\Delta)$. The trapping parameters are $\bm{\alpha}=diag(
\alpha_x, \alpha_y, \alpha_\phi )$, and $\beta$ and $\gamma$ are the controller weights that are tuned during the experiment to obtain the desired performance. Selection of $\alpha_x$, $\alpha_y$, and $\alpha_\phi$ gives the user flexibility in choosing the level of control authority over each individual state (2D COM position $(x,y)$ and orientation $\phi$). Following the MPC strategy, we minimize the objective function in Eq. \ref{eq:singleparticlecontrol} at each sampling instant to obtain the flow rates over the entire horizon, but we only apply the flow rates corresponding to the first MPC interval by setting $\bm{q}(t_k) = \tilde{\bm{q}}(t_k)$ and resample the positional and orientational state of the particle $\bm{X}$ at the next sampling instant (Supplemental Material and Fig. S3, Fig. S4). For experimentally implementing MPC, we use the toolkit for Automatic Control and Dynamic Optimization (ACADO) \cite{houska_acado_2011,quirynen_autogenerating_2015}.

\subsection{Trajectory tracking and control}
In this work, we implement nonlinear model predictive path-following control to precisely manipulate the COM position of colloidal particles along user-defined trajectories \cite{faulwasser_implementation_2016}. Nonlinear path-following control is implemented by parameterizing the reference trajectory $\bm{r}$ using a parameter $\theta$ such that $\bm{r}(\theta):[-1,0]\mapsto{\mathbb{R}^2}$ represents the desired trajectory. The MPC formulation in Eq. \ref{eq:singleparticlecontrol} is then modified as follows:
\begin{subequations}\label{eq:trajectorytracking}
	\begin{align}
	\underset{\tilde{x},\tilde{q},\theta}{\min }\ J &=\sum^{t_k+T_N-1}_{\tau=t_k} \bigl\{  \| \alpha\left(\tilde{\bm{x}}(\tau)-\bm{r}(\tilde{\theta})\right) \|^2 +  \delta_1 \|\tilde{\theta}(\tau)\|^2 \nonumber\\
	&+ \beta \|\tilde{\bm{q}}(\tau)\|^2 +\delta_2 \|\tilde{\Phi}(\tau)\|^2 \bigr\} \nonumber\\
	&+ \gamma \|\tilde{\bm{x}}(t_k+T_N)-\bm{r}(\tilde{\theta}(t_k+T_m))\|^2 \\
	\text{s.t.} \ &\frac{d \tilde{\bm{x}}}{dt}= \bm{F}(\tilde{\bm{x}},\tilde{\bm{q}},\bm{R}),\quad \tilde{\bm{x}}(t_k) = \bm{x}(t_k)\\
	& \dot{\tilde{\theta}} = -\lambda\tilde{\theta}+\Phi_{max}-\tilde{\Phi},\quad \tilde{\theta}(t_k) = \theta(t_k|\theta(t_{k-1})) \label{eq:controls_timinglaw}\\
	& 0\leq \tilde{\Phi}(\tau) \leq \Phi_{max}\\
	&\sum^N_{i=1}   \tilde{q}_i(\tau) =0 \quad \forall \ \tau = t_k,\ldots, t_k+T_N
	\end{align}	
\end{subequations}
Eq. \ref{eq:controls_timinglaw} is known as the timing law because it controls the evolution of the path parameter $\theta$. $\lambda$ is a small value ($\approx$0.001) that is used to stabilize the timing law, $\Phi$ is the speed of the set point along the reference trajectory, and $\Phi_{max}$ denotes the maximum permissible value of $\dot{\theta}$. $\Phi_{max}$ can be tuned based on the rate required to track the reference trajectory. At each sampling instance, once the current positional state $\bm{x}$ of a particle is known, this information is projected onto the reference trajectory to determine the nearest point on the trajectory, which ultimately enables determination of the corresponding $\tilde{\theta}$. However, estimating $\tilde{\theta}$ in real time such that the distance between the actual state and reference state is minimized at each sampling instant is non-trivial and corresponds to a non-convex problem. To address this issue, at each sampling instance, the initial condition for $\tilde{\theta}$ is set equal to the predicted value of $\theta$ from the previous sampling instance. At the beginning of the path, $\theta$ is set to $-1$ in the timing law for the first sampling instance.

Using this approach, we effectively control the COM position of colloidal particles along precisely defined paths. Prior work has only considered fluidic manipulation of the COM position of multiple isotropic particles \cite{shenoy_stokes_2016}, where positional control was simply achieved by merely stepping the set-point positions along a pre-determined path at a constant speed. Unfortunately, this simplistic approach does not consider the actual particle position during the event, which often results in trapped particles migrating far away from the set point during the stepping process, resulting in large deviations from the reference trajectory. The path-following trajectory control method implemented in the present work overcomes these limitations by coordinating the set-point motion with the motion of the particle, such that the set-point speed along the trajectory can be increased or decreased based on the lag distance of the particle. 

\subsection{Set-point control for precise manipulation of particle orientation}
We began by manipulating the 2D COM position and orientation of single anisotropic Brownian particles with simultaneous control. In particular, we trapped and controlled the orientation of rigid, rod-like polystyrene particles (Fig. \ref{fig:trap_stiff}; Supplementary Movie 1). In this experiment, a target particle is initially oriented at an angle of $10^{\circ}$ with respect to the inflow axis. At $t=0$, the desired set point for particle orientation is changed to $80^{\circ}$. The controller generates a set of flow rates to rotate the particle, and the desired positional and orientation state is reached in $\approx 6.7$ s. We note that the ends of the rod-like particles are indistinguishable, so the orientations $\phi$ and $(180^{\circ}-\phi)$ are equivalent. To quantitatively analyze trap performance, we confined a single anisotropic rod at a desired orientation of $\phi = 80^{\circ}$ for $\approx 200$ s while tracking the COM position and orientation (Fig. \ref{fig:trap_stiff}a and Fig. \ref{fig:trap_stiff}b). The probability distributions of orientation and COM position are shown in Fig. \ref{fig:trap_stiff}a and Fig. S5, which are well described by Gaussian distributions with standard deviations of translational displacement $\sigma_x=0.55$ $\mu$m and $\sigma_y=0.20$ $\mu$m, and a standard deviation of angular displacement $\sigma_{\phi}=2.1^{\circ}$. Standard deviations of translational and orientational displacement provide a measure of the trap tightness-of-confinement, which in the case of the translational displacement is much smaller than the major axis of the particle.

\begin{figure}
	\small
	\centering	
	\includegraphics[width=0.5\textwidth]{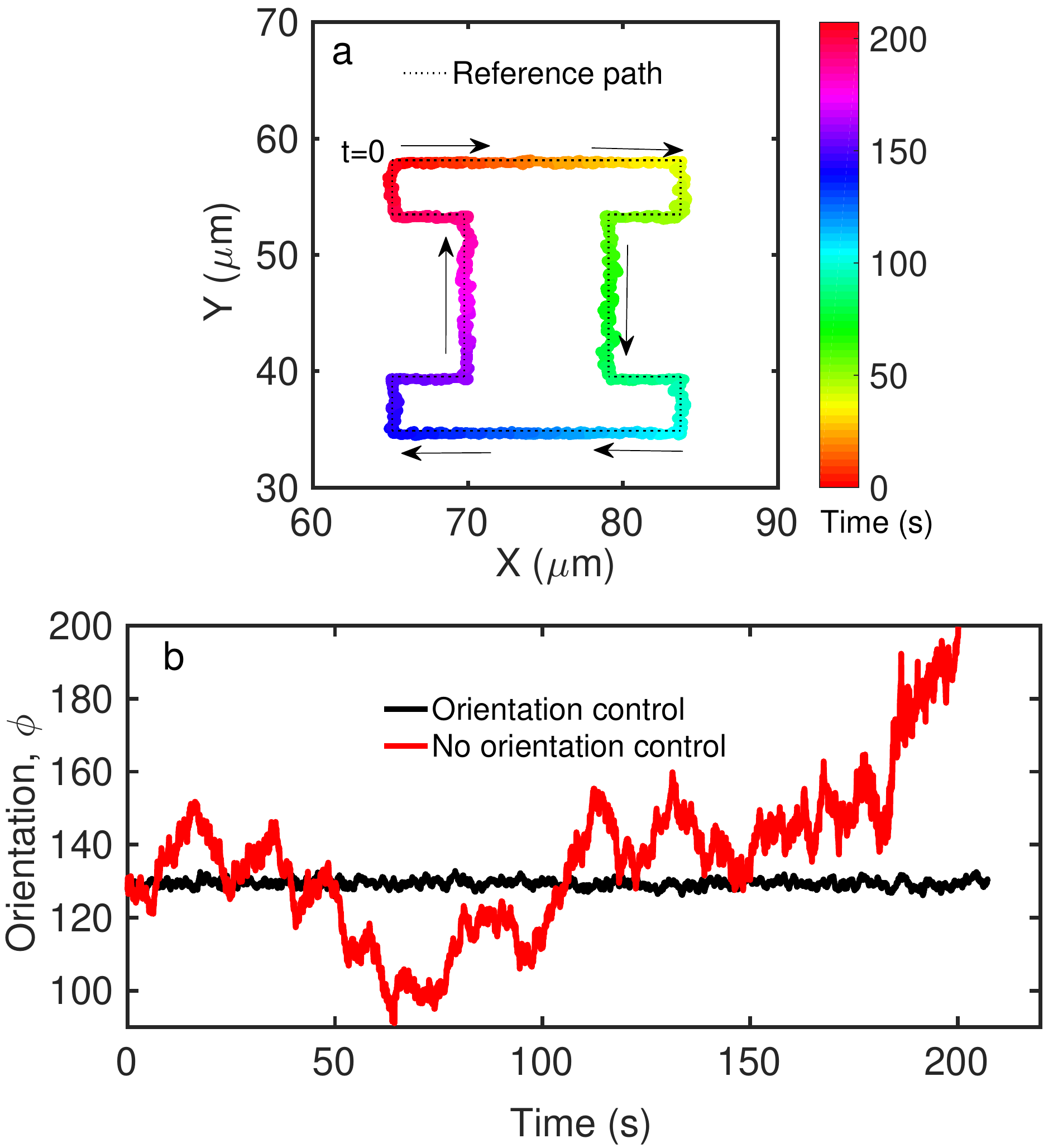}
	\caption{\label{fig:trap_I} Simultaneous manipulation of 2D center-of-mass position and orientation angle for single anisotropic particles using fluid flow. (a) The position of a rod-like Brownian particle was controlled to trace the letter `I' while maintaining a constant orientation angle of $\phi=130^{\circ}$ throughout the path. The actual trajectory of particle is shown in colormap, while the reference path is shown in black dotted markers. (b) Comparison between the active control and no control over rod orientation angle during the 2D center-of-mass manipulation experiment.}
\end{figure}

We further characterized the translational and angular trapping stiffness by determining the power spectral density (PSD) of positional and angular fluctuations of trapped Brownian rods (Fig. \ref{fig:trap_stiff}b) \cite{leal_advanced_2007,kang_angular_2012}. Trap stiffness $\kappa$ is determined from the corner frequency $f_c$ of the PSD by fitting the experimentally determined power spectrum to a Lorentzian function using the Levenberg-Marquardt algorithm, such that $\kappa_i=2\pi\zeta_i f_{ci}$ where $\zeta_i$ is the hydrodynamic friction coefficient and $i=x,y,\phi$ represents the value of the parameter along the directions $x$, $y$, and $\phi$, respectively. In this way, we determined trap stiffness values to be $\kappa_x=8.3 \times 10^{-4}$ pN/nm, $\kappa_y=2.7 \times 10^{-3}$ pN/nm, and $\kappa_{\phi}=9.1 \times 10^{3}$ pN$\cdot$nm/rad, respectively. The values of translational trap stiffness $\kappa_{x,y}$ are comparable to a weak optical trap \cite{neuman_single-molecule_2008} and to a Stokes trap for confining $2.2$ $\mu$m diameter spherical particles \cite{shenoy_stokes_2016}. Prior work based on optical traps reported an angular trap stiffness as $\kappa_{\phi} = 3.36 \times 10^{3}$ pN$\cdot$nm/rad for 1 $\mu$m diameter quartz particles \cite{la_porta_optical_2004}, which is similar to the angular trap stiffness determined using our flow-based approach.

We further demonstrated the ability for controlled orientational manipulation by rotating a rod-like particle from an initial orientation angle $\phi=120^{\circ}$ to a final angle $\phi=45^{\circ}$ (Supplementary Movie 2) in a four-channel microfluidic device. In this experiment, the initial and final values of orientations are located in different quadrants in the cross-slot channel (see Fig. \ref{fig:rod_device}b). While initially approaching the target orientation, the flow tends to orient the rod along extensional axis ($90^{\circ}$ and $270^{\circ}$ in our convention from Fig. \ref{fig:rod_device}b). However, trapped particles aligned in these orientations occasionally receive a Brownian kick which moves the orientation angle to $\phi<90^{\circ}$, after which the particle quickly reaches the target orientation. The trajectory for the rod orientation reaching the target angle reveals an interesting feature of using an $M=4$ channel device for controlling anisotropic rods. Here, it is technically possible to control 3 independent degrees of freedom ($x,y$ position and orientation angle $\phi$), but there are no additional degrees of freedom to aid in controller flexibility. A rod oriented along the extensional axis at $90^{\circ}$ represents a stable conformation in flow, and a Brownian fluctuation to yield $\phi<90^{\circ}$ is then leveraged to reach the final desired angle of $\phi=45^{\circ}$. We performed numerical simulations for non-Brownian particles, and indeed observed that the rod never leaves the stable conformation position $\phi=90^{\circ}$ and hence, the desired orientation is not achieved. In experiments, however, the rotational Brownian kick effectively pushes a trapped rod out of the stable orientation and into the desired quadrant of the final orientation. To enhance orientation control, we also used microfluidic device with $M=5$ channels for controlling anisotropic particles with 3 degrees of freedom, which enabled facile trapping and control of orientation angle over the entire $2\pi$ range without needing to rely on Brownian motion to overcome stable orientation alignments in flow (Supplementary Movie 3). In this way, simply using microdevices with at least one additional channel relative to the number of degrees of freedom loosens the constraints and improves orientational controllability. 

\subsection{Set-point control for simultaneous manipulation of particle position and orientation}
Individual rod-like particles can be also manipulated in two dimensions along specific desired trajectories, both in positional and orientational space (Fig. \ref{fig:trap_I}). In this way, anisotropic particles can be controlled to maintain a constant orientation angle $\phi$ during 2D COM positional control, or alternatively to follow a specific orientation `program' for $\phi$ while tracking a defined trajectory during a manipulation event. To illustrate this level of control, the 2D COM position of a particle is steered to trace the capital letter `I' while maintaining a constant orientation angle $\phi=130^{\circ}$ throughout the path, as shown in Fig. \ref{fig:trap_I}a (Supplementary Movie 4). Here, the trajectory is discretized into many segments for which the target COM position is slowly and repeatedly moved along the letter I. In this manner, the particle covers a predetermined trajectory over a relatively large distance (120 $\mu$m) with high spatial and angular accuracy over $\approx$200 s. In Fig. \ref{fig:trap_I} a, we show the 2D COM positions of a rod in color-coded trajectory overlaid with the actual positions of the set path in black dashed markers. We observe that the rod follows the reference path closely with small deviations. For comparison, we controlled a rod to move along the same path without actively controlling the orientation of the particle, and the results show that the orientation angle $\phi$ fluctuates significantly during the manipulation event (Fig. \ref{fig:trap_I}b). We anticipate that particle trajectory deviations will increase if the speed of movement along the reference path is increased (Supplemental Material and Fig. S6). 

\begin{figure}
	\small
	\centering	
	\includegraphics[width=0.5\textwidth]{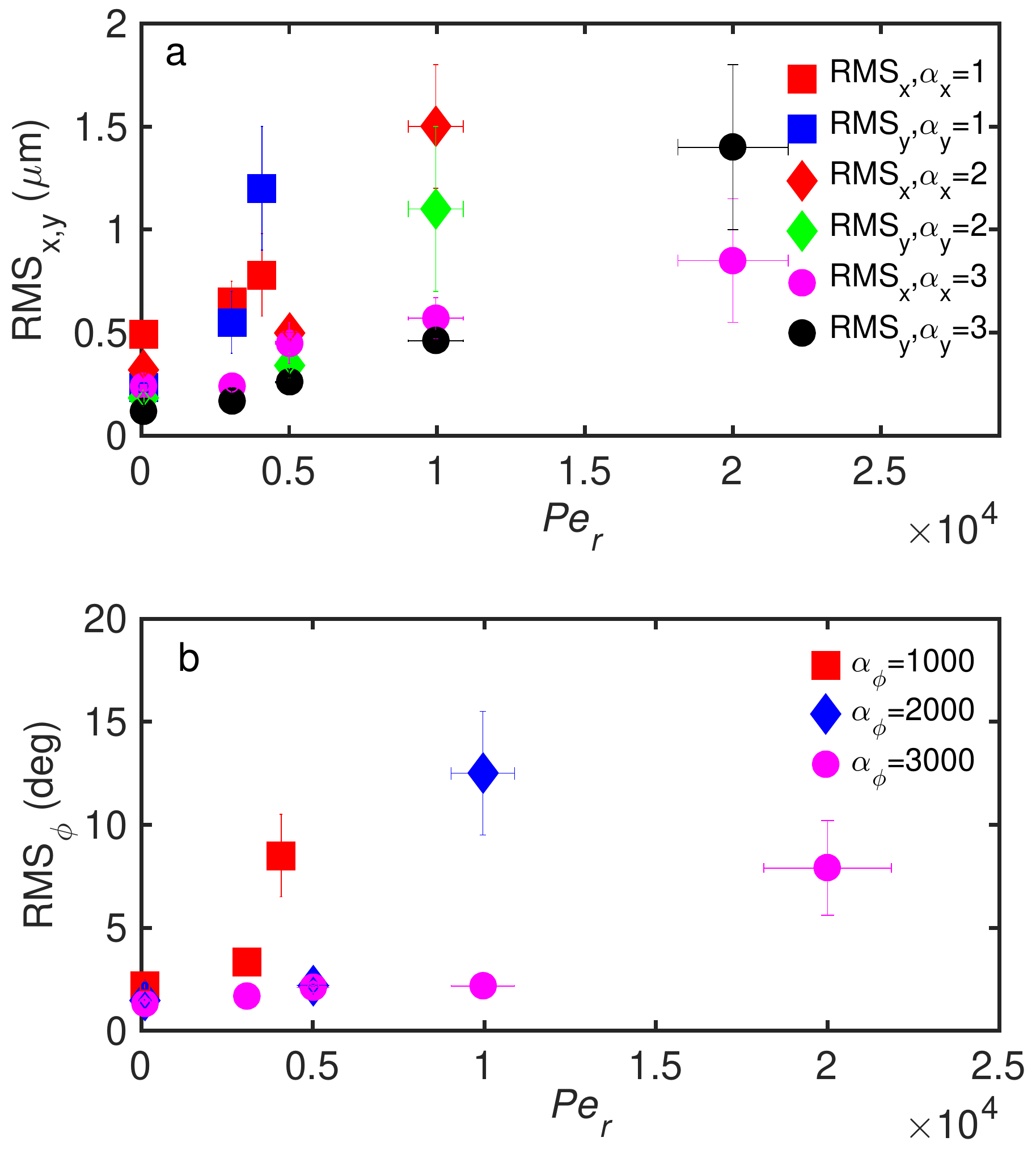}
	\caption{\label{fig:trap_rms} Characterization of trap response as a function of rotational Peclet number $Pe_r$. (a) Tightness of confinement along the translational directions as a function of $Pe_r$ for different values of trapping parameters: $RMS_x$ (red squares), $RMS_y$ (blue squares) at $\alpha_x=1, \alpha_y=1, \alpha_{\phi}=1000$, $RMS_x$ (red diamonds), $RMS_y$ (green diamonds) at $\alpha_x=2, \alpha_x=2, \alpha_{\phi}=2000$, and $RMS_x$ (magenta circles), $RMS_x$ (black circles) at $\alpha_x=3, \alpha_y=3, \alpha_{\phi}=3000$. (b) Tightness of confinement $RMS_{\phi}$ along the angular direction at varying $Pe_r$ for different values of trapping parameters: (red squares) at $\alpha_x=1, \alpha_y=1, \alpha_{\phi}=1000$, (blue diamond) at $\alpha_x=2, \alpha_y=2, \alpha_{\phi}=2000$, (magenta circles) at $\alpha_x=3, \alpha_y=3, \alpha_{\phi}=3000$.}
\end{figure}

\begin{figure}[t]
	\begin{center}
		\includegraphics[width=0.5\textwidth]{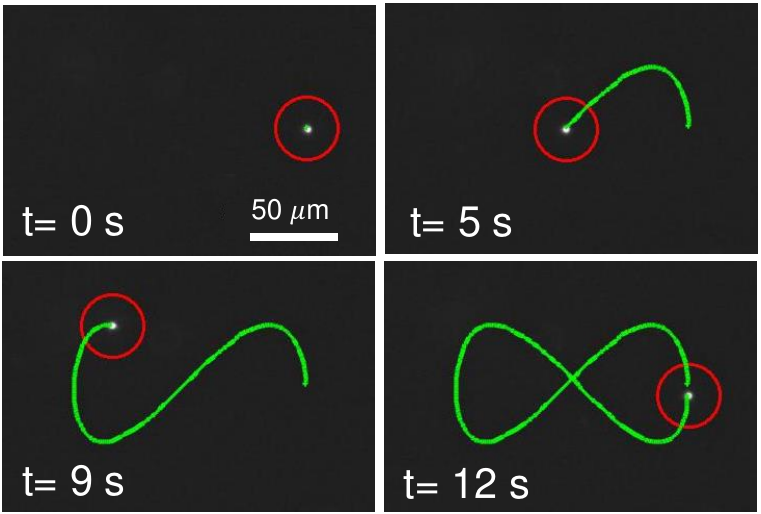}
		\caption{\label{fig:figure8traj} Trajectory path control for manipulating a spherical Brownian particle (2.2 $\mu$m diameter bead) over a `figure-8' curve. The positional history of the particle is shown with a green line. MPC control parameters were set to $\alpha_x=1, \alpha_y=1$, $\alpha_{\phi}=0$, $\lambda=0.001$ and $\Phi_{max} = 0.05$ for the these experiments.}
	\end{center}
\end{figure}

Using the flow-based trapping method reported in this work, anisotropic particles can be maintained at a desired state under no-flow conditions or under finite flow (net-flow) conditions. Under no-flow conditions, the controller applies a small magnitude fluid flow to correct for minor fluctuations due to rotational or translational Brownian motion of a particle. For trapping anisotropic particles in $M=4$ channel devices using net-flow conditions, we observed that anisotropic particles can be confined at a desired orientation, but only up to a critical value of fluid strain rate $\dot{\epsilon}$. For very high strain rates, anisotropic particles tend to align along the extensional flow axis. These observations suggest that the fluid strain rate $\dot{\epsilon}$ (Supplemental Material and Fig. S7) plays a key role in determining trap performance and stability. To investigate this further, we characterized the effect of rotational Peclet number $Pe_r = \dot{\epsilon}/D_r$ on particle confinement (Fig. \ref{fig:trap_rms}a,b), where $D_r$ is the rotational diffusion coefficient of an anisotropic particle (Supplemental Material and Fig. S8). Here, we performed a series of experiments by varying $Pe_r$ for a trapped particle while controlling the orientation angle at $\phi=130^{\circ}$. $Pe_r$ is varied by changing $\delta P$, which is the pressure difference between inlet channels 1,3 and outlet channels 2,4 in Fig. \ref{fig:rod_device}b. Initially, all four channels were maintained at a base pressure $P_0 =2.5$ psi. For applying a net $\delta P$, the inlet channels 1,3 were maintained at a higher pressure than the base value $P_0$. The strain rate $\dot{\epsilon}$ is determined by particle tracking microscopy (Supplemental Material and Fig. S7). 

Using this approach, we determined the root-mean-square (RMS) translational $(RMS_x, RMS_y)$ and angular displacements $(RMS_{\phi})$ of a trapped particle as a function of $Pe_r$ and the controller parameters $\alpha_x$ and $\alpha_y$ (Fig. \ref{fig:trap_rms}a,b). In all cases, the RMS translational displacements are $<$1.5 $\mu$m over a wide range of $Pe_r$, suggesting that the trap is robust up to at least $Pe_r = 20,000$. In addition, the translational displacements $RMS_{x,y}$ of anisotropic particles increase as $Pe_r$ increases. Interestingly, this observation is in contrast to the behavior of a spherical particle trapped in a Stokes trap \cite{shenoy_stokes_2016}, where the RMS particle displacement increases (decreases) as the $Pe_r$ increases along the extensional (compressional) axis. These results reflect the performance of the trap given the added constraint of controlling particle orientation in addition to 2D position for anisotropic particles. 

\begin{figure}[b]
	\begin{center}
		\includegraphics[width=0.5\textwidth]{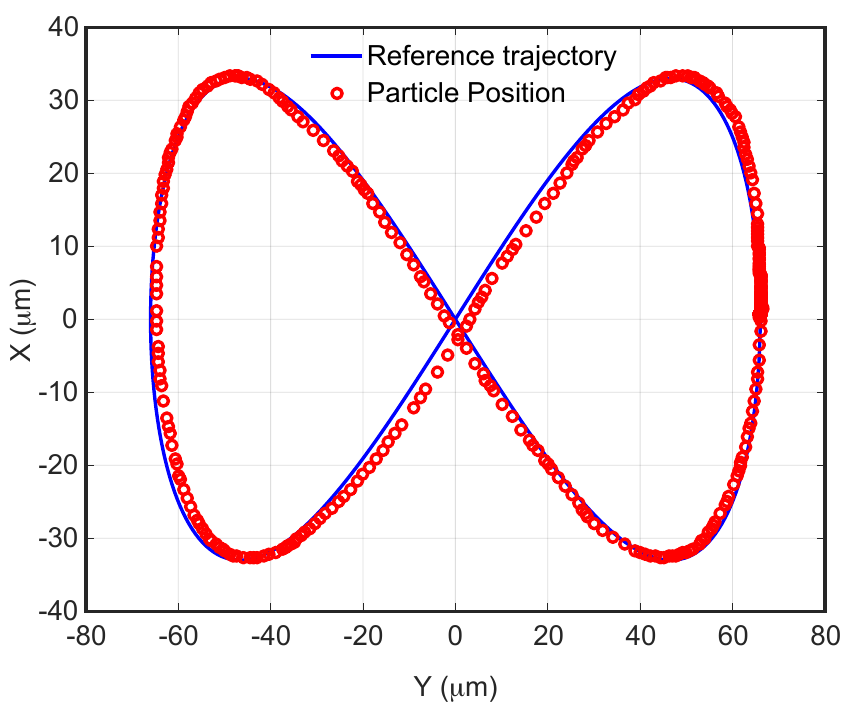}
		\caption{\label{fig:controls_traj_xy} Actual and reference paths for a spherical Brownian particle tracing a `figure-8' curve. The reference trajectory is shown in blue, with the red circles marking the particle position over the course of the trajectory.}
	\end{center}
\end{figure}

Fig. \ref{fig:trap_rms}b shows the RMS rotational displacements as a function of $Pe_r$ and the controller parameter $\alpha_{\phi}$. For small values of $\alpha_{\phi}$, the angular trapping displacement $RMS_{\phi}$ increases as the value of $Pe_r$ increases, such that the flow tends to align the rod along the extensional axis at a critical value of $Pe_r$, denoted by the last data point shown for each value of $\alpha_{\phi}$ in Fig. \ref{fig:trap_rms}b. Interestingly, by tuning the trapping parameter $\alpha_{\phi}$, we are able to increase the tightness of confinement and achieve the desired orientation over a wide range of $Pe_r$. In this manner, we confirmed that the tightness of confinement along both the translational and angular directions can be tuned in real-time by proper adjustment of trapping parameters. The effect of trapping anisotropy, which refers to different trapping stiffnesses in the $x,y$-directions, can be further quantified (Supplemental Material and Fig. S9 a,b).

In addition to simultaneously controlling the position and orientation of a single anisotropic particle with high precision, this trapping method described by Eq. \ref{eq:singleparticlecontrol} enables simultaneous translational and rotational control of {\em multiple} anisotropic particles. This feature is demonstrated using numerical simulations (Supplemental Material, Supplementary Movie 5), where the translational and rotational motion of two anisotropic particles was independently controlled in a microdevice corresponding to $M=7$ channels. 

\begin{figure}
	\begin{center}
		\includegraphics[width=0.5\textwidth]{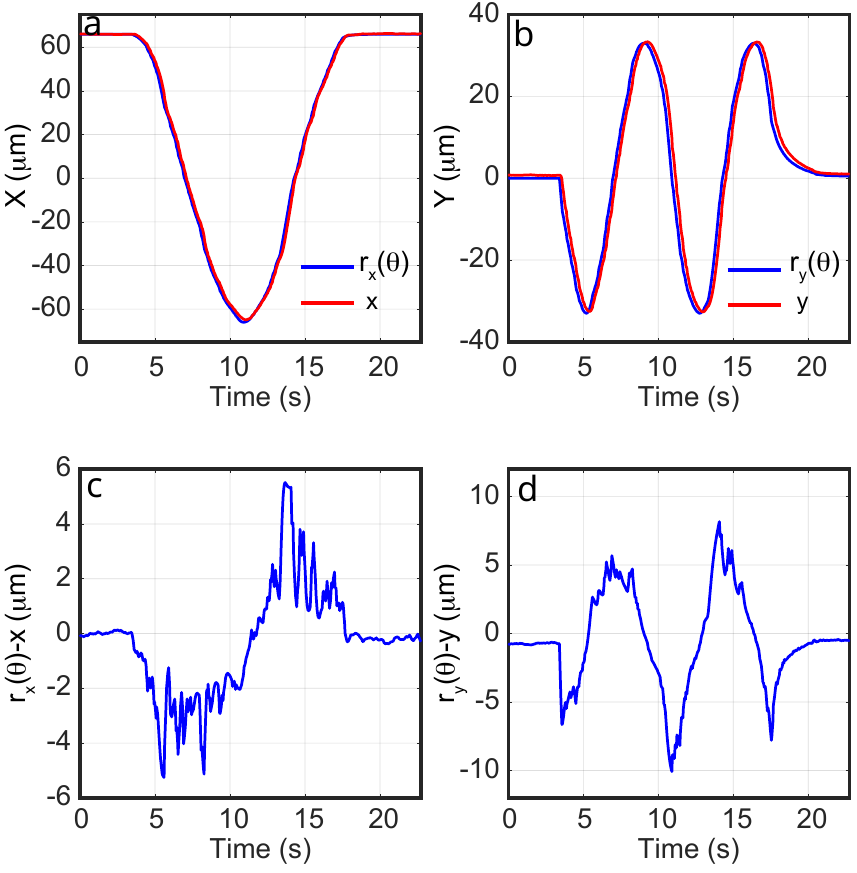}
		\caption{\label{fig:controls_traj_xyerror} Reference path and the actual path of an isotropic particle during trajectory path control. (a), (b) $x$ and $y$ coordinates of the particle and the reference trajectory as a function of time, respectively. (c) Offset error between the $x$ coordinate of the particle and the $x$ coordinate of the reference trajectory as a function of time. (d) Offset error between the $y$ coordinate of the particle and the $y$ coordinate of the reference trajectory as a function of time.}
	\end{center}
\end{figure}

\subsection{Trajectory control of isotropic Brownian particles}
We first implemented trajectory path control by manipulating the 2D COM position of a spherical Brownian particle (2.2 $\mu$m diameter fluorescent bead). Particle position was controlled to follow a figure-8 (Fig. \ref{fig:figure8traj} and Supplementary Movie 6), where the trajectory is described by a parametric equation such that $\bm{r}(\theta) = [30\cos(-2\pi\theta),\ 15\sin(-4\pi\theta)]^T$, with $\theta\in[-1,0]$. Fig. \ref{fig:controls_traj_xy} shows the reference trajectory overlaid with the actual trajectory of the particle. We observe that the particle closely follows the reference trajectory with only small deviations from the reference trajectory. The time-dependent particle position coordinates $x$ and $y$ during the experiment are shown in Fig. \ref{fig:controls_traj_xyerror}a,b, which further illustrates that the particle closely tracks the reference trajectory with only minor deviations. Importantly, the particle covers a large distance of approximately 100 $\mu$m over a short period of time ($\approx$12 s). We emphasize that the trajectory control approach provides a dramatic improvement over the prior approach of simply stepping the set-point position at a fixed speed along a desired trajectory, which generally requires longer times and results in larger deviations from reference trajectories during manipulation events ({\em e.g.} requiring 300 s to manipulate a bead over a  200 $\mu$m distance) \cite{shenoy_stokes_2016}.

\begin{figure*}
	\begin{center}
		\includegraphics[width=1.0\textwidth]{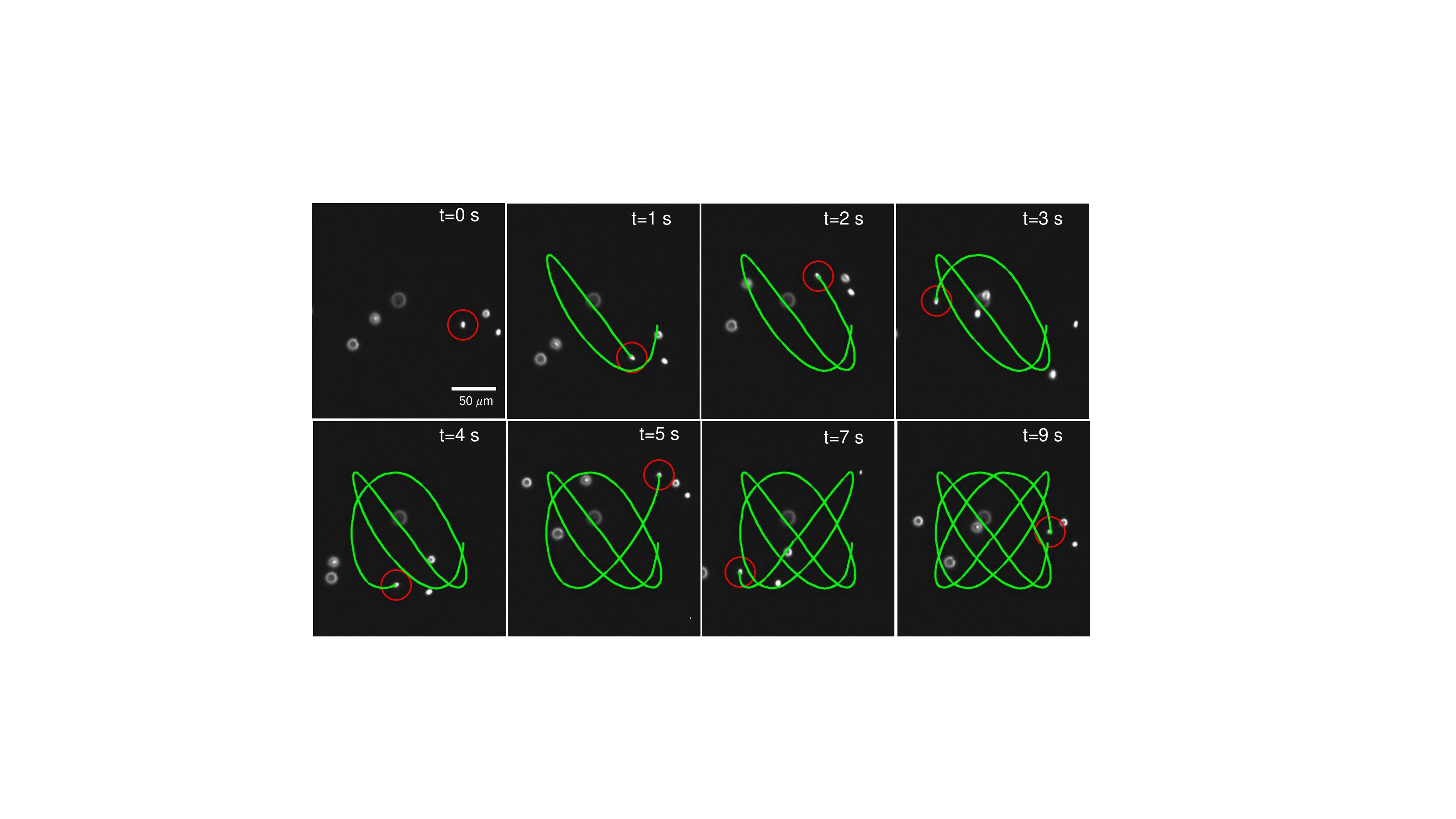}
		\caption{\label{fig:lissajous} Trajectory path control for manipulating an isotropic Brownian particle (2.2 $\mu$m diameter spherical bead) over a complex parametric curve. The positional history of the particle is shown with the green line. The particle closely follows a complex parametric path and traverses a linear distance of several hundred microns in only $\approx$9 s.}
	\end{center}
\end{figure*}

The offset error between the desired set point (corresponding to the current value of the parameter $\theta$) and the actual instantaneous particle position is shown in Fig.\ref{fig:controls_traj_xyerror}c,d. Interestingly, it is observed that the offset error generally increases during the straight sections of the reference trajectory and decreases when the set point occurs in regions of high curvature. In brief, the set-point stepping motion slows down when approaching regions of high curvature, thereby allowing the particle to catch up to the set-point motion. The offset errors shown in Fig. \ref{fig:controls_traj_xyerror}c,d correspond to the difference between the set point position and the particle's actual position, but not the distance between the particle's position and its projection onto the reference trajectory. Thus, even though offset errors show an approximate value of $\approx$5-10 $\mu$m, the projected errors are significantly smaller because the particle often lags behind the set-point but remains on the curve. Offset errors can be reduced further by appropriately tuning the weights corresponding to the difference between the set-point and the particle's position and the weights corresponding to the flow rates $\bm{q}$. 

The ability of trajectory path control to achieve robust manipulation at relatively fast speeds is further demonstrated by moving a spherical bead along more complicated parametric curves, as shown in Fig. \ref{fig:lissajous} and Supplementary Movie 7. Here, the particle effectively follows a complex path and rapidly traverses a linear distance of several hundred microns in only $\approx$9 s. In addition to trajectory tracking control in position space, this method also enables trajectory control of anisotropic particles in both position and orientation space. This feature is demonstrated using numerical simulations (Supplemental Material, Supplementary Movie 8), where an anisotropic particle is moved along a parametric curve described by a figure-8 while simultaneously changing its orientation from $89^{\circ}$ at the beginning of the path to $1^{\circ}$ midway, and back to $89^{\circ}$ at the end of reference path. In this way, we are able to achieve precise trajectory control over both the position and orientation of anisotropic particles by simply adding additional state variables in the MPC objective function Eq.\ref{eq:trajectorytracking}. These results demonstrate the robustness and scalability of our approach, where we can steer the particles along a reference path in an output space (position and orientation) without needing to fix the speed of movement along the path beforehand. 

\subsection{Transient and steady-state orientation dynamics of Brownian rods in extensional flow}
We used the new hydrodynamic trap method to directly observe the transient and steady-state orientation dynamics of anisotropic Brownian particles in extensional flow. For these experiments, the 2D particle COM is confined, but the particle orientation is not controlled and hence governed by flow-dependent dynamics. Orientation trajectories are recorded as a function of $Pe_r$. A trajectory showing the transient orientation dynamics for a single anisotropic Brownian rod at $Pe_r=602$ is shown in Fig. S10 (Supplemental Material). 

\begin{figure*}
	\begin{center}
		\includegraphics[width=1.0\textwidth]{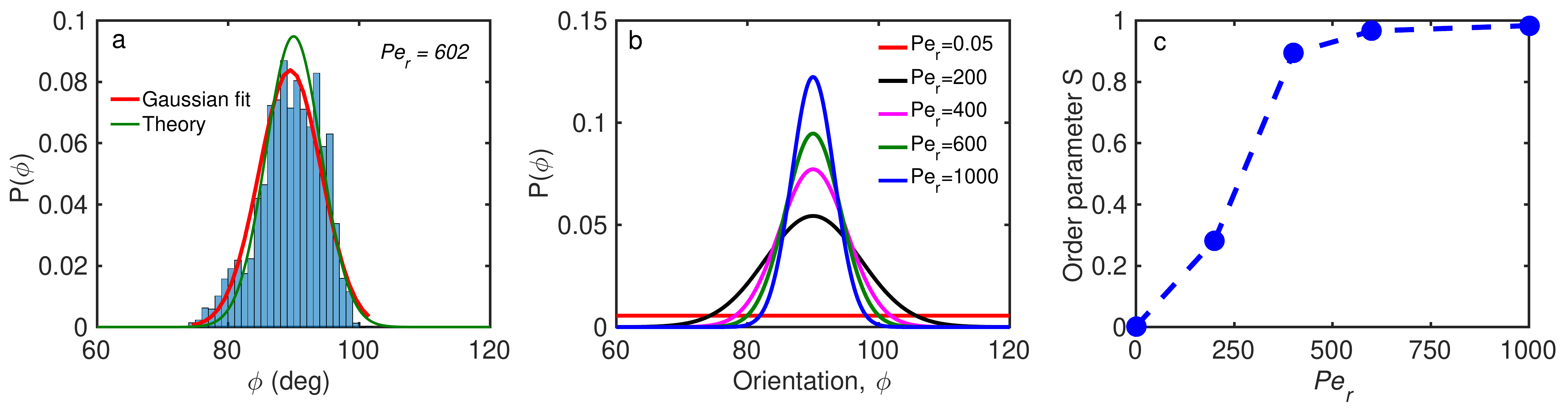}
		\caption{\label{fig:order_parameter} Orientation dynamics of a confined anisotropic particle. (a) Comparison of probability distribution function from experiments, and theory. (b) Steady-state distribution of the anisotropic particle at different Peclet number from theory. (c) Variation of order parameter as a function of Peclet number from theory.}
	\end{center}
\end{figure*}

To complement experiments, we determined the probability distribution of Brownian rod orientation in extensional flow using analytical model. The full orientation distribution function (PDF) $\psi(\phi,t)$ of a Brownian particle is described by the Fokker-Planck equation (FPE) \cite{doi_theory_1988}:
\begin{equation}\label{eq:FPE}
\frac{\partial \psi\left ( \phi, t \right )}{\partial t}=D_r\frac{\partial^2 \psi\left ( \phi, t \right ) }{\partial \phi^2}-\frac{\partial \left [\dot{\phi}    \psi\left ( \phi, t \right ) \right ]}{\partial \phi}
\end{equation}
The steady-state distribution has a familiar Boltzmann form (Supplemental Material):
\begin{equation}\label{eq:FPE_soln}
\psi\left ( \phi \right )=\frac{e^{-\frac{q_1+q_3-q_2-q_4}{\pi HR^2D_r}\cos2\phi}}{\int_{0}^{\pi}e^{-\frac{q_1+q_3-q_2-q_4}{\pi HR^2D_r}\cos2\phi}}
\end{equation}
where $q_1, q_2, q_3, q_4$ are the flow rates through the channels of microfluidic device, $H$ is the depth of device, and $R$ is the width of cross-slot channel.

The experimentally determined orientation distribution function agrees well with the analytical model given by Eq. \ref{eq:FPE_soln} (Fig. \ref{fig:order_parameter}a). As $Pe_r$ increases, the orientation PDF becomes sharply peaked along the axis of extension, as shown in Fig. \ref{fig:order_parameter}b. To characterize the degree of alignment along the extensional axis, we also define a 2D order parameter as the eigenvalue of tensor $\bm{S}=2\left< \bm{p}\bm{p}\right> - \boldsymbol{\delta}$, where $\bm{p}$ is the unit vector along the major axis of the particle such that $\bm{p} = [\cos \phi, \sin \phi]^{T} $. In the limit of $Pe_r\rightarrow\ 0$, the order parameter $S$ tends to zero, and the distribution becomes isotropic. For large values of $Pe_r$, the order parameter $S$ approaches unity and the distribution is strongly aligned along the extensional axis (Fig. \ref{fig:order_parameter}c).

\begin{figure}
	\begin{center}
		\includegraphics[width=0.5\textwidth]{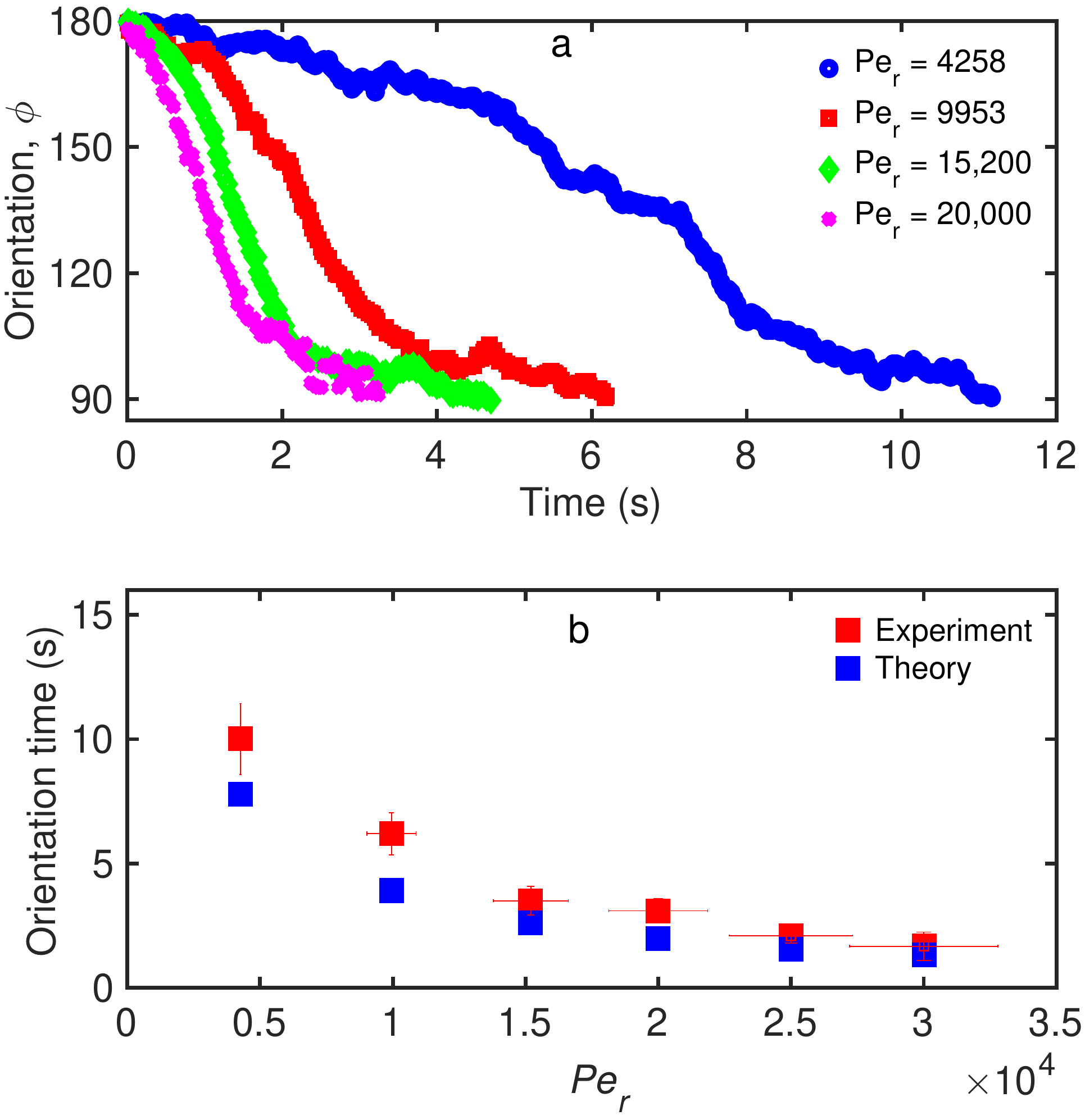}
		\caption{\label{fig:timetoorient} Transient orientation dynamics of an anisotropic Brownian rod in extensional flow. (a) Transient orientation angle $\phi$ of a trapped rod during the reorientation process. (b) Characteristic reorientation time for a single rod following a step change in orientation angle set-point. }
	\end{center}
\end{figure}

In addition to the orientation PDF and order parameter $S$, we further determined the characteristic time required by a Brownian rod to reorient in flow following a step change from an initial orientation angle set point to different set point. In order to quantitatively characterize the reorientation time as a function of flow strength, we record the transient trajectories of a single rod following the step change of orientation from $\phi=180^{\circ}$ to $\phi=90^{\circ}$ as a function of $Pe_r$ (Fig. \ref{fig:timetoorient}a). The average reorientation times are plotted in Fig. \ref{fig:timetoorient}b. Our results show that the reorientation time decreases upon increasing $Pe_r$, which is in agreement with a simple analytical model (Supplemental Material). Such results can be extremely useful for describing the non-equilibrium phase diagram of anisotropic rods in extensional flow, which can be used to describe the transition from an isotropic (disordered) state to an aligned (ordered) state in suspensions of dilute rigid rods or liquid crystals as a function of flow strength in time-dependent flows. 

\section{Discussion}
In this work, we develop and implement fundamentally new flow-based methods for particle manipulation that allow for simultaneous control over the 2D position and orientation of anisotropic Brownian particles. Moreover, these methods enable path-following control for manipulating colloidal particles along defined reference trajectories, where the speed of movement is a degree of freedom in the controller design. Trap performance is characterized by translational and angular trapping stiffness, and these results show that this method can be used to stably control anisotropic particles over a wide range of flow strengths (rotational Peclet number $Pe_r$). Moreover, this technique can be used for fluidic manipulation of three planar degrees of freedom ($x, y, \phi$) of anisotropic particles. Future application of this method to trapping multiple particles holds the potential to enable the study of interparticle interactions as a function of relative spacing, particle orientation, and imposed flow strength. 

Unlike alternative techniques that exploit intrinsic material properties to manipulate particles ({\em e.g.} index of refraction, magnetic properties, surface charge), flow-based methods impose no restrictions on the physical or chemical properties of trapped particles, and hence can be used for small particles of any material and size, given that they can be imaged or detected. In the current setup, we only control the 2D COM position and orientation in the $x-y$ plane, without control in the orthogonal $z$-direction. From this view, the current method is not directly amenable for control over particle position in the $z$-direction or out-of-plane rotation. Nevertheless, we generally find that out-of-plane particle orientation is rare and only occurs over extremely long timescales (minutes to hours) at which passive Brownian motion may lead to orientation changes. Overall, this flow-based trapping method provides stable and robust control of the orientation and position of anisotropic particles.

A large number of prior of studies in colloidal science and biophysics have relied on trapping a small particle at a fixed set-point by suppressing disturbances using optical \cite{grier_revolution_2003}, magnetic \cite{winkleman_magnetic_2004}, electric \cite{cohen_suppressing_2006}, or fluidic forces \cite{schroeder_single_2018}. In this work, we experimentally demonstrate a path-following control method to manipulate particles along arbitrary reference trajectories, where the speed of motion along the path is also a state variable in the controller design. Our results show that particles can be manipulated across complex paths and trajectories by changing only a few parameters in the MPC objective function, thereby providing a level of control that has not been previously possible using conventional set-point stabilization trapping techniques. Our work effectively extends the implementation of path-following control to millisecond time scale system dynamics. Taken together, these manipulation methods can be readily applied to various problems where the speed of movement of particles along a desired path is not fixed {\em a priori}, for example in systematically building higher-order and complex assemblies of structurally and chemically different anisotropic particles. Moving forward, flow-based trapping and manipulation techniques hold the potential to benefit fundamental studies in multiple fields including soft materials, colloidal science, and biophysics.

\section*{Acknowledgements}
This work was supported by the National Science Foundation by Award \# NSF CBET 1704668. A.S acknowledges support from FMC for an Educational Fund Fellowship.

\section*{Author contributions}
  D.K.,A.S., C.M.S. conceived the project and designed the experiments; D.K. and A.S. performed the experiments and analyzed the data; S.L. synthesized the rod-like particles; D.K, A.S., C.M.S. wrote the paper; C.M.S supervised the research. All authors agreed on the final manuscript.

\end{document}